\documentclass[aps,prc,reprint,amssymb,amsmath,superscriptaddress, floatfix,10pt]{revtex4-1}
\usepackage{mathptmx}
\usepackage{graphicx}
\usepackage{color}  
\makeindex
\newcommand{\textblue}{\textcolor[rgb]{0.00,0.07,1.00}}

\begin{document}

\title{The mean-field theory of superfluid-superconducting vortex states in the outer core of neutron stars}

\author{Dmitry Kobyakov}
\email{dmitry.kobyakov@appl.sci-nnov.ru}
\date{May 10, 2026}
\affiliation{Institute of Applied Physics of the Russian Academy of Sciences, 603950 Nizhny Novgorod, Russia}

\begin{abstract}
\textbf{Background:} In the outer core of neutron stars, a quantum vortex state is expected to exist reminiscent of superfluid vortex and superconducting fluxtube. The pairing critical temperatures $T_{c\alpha}$, $\alpha=\{n,p\}$ are pressure-dependent extending in a macroscopic region, giving rise to inhomogeneity of the superfluid mean-field in the isothermal outer core. The earlier mean-field theory either covered only the special case $T_{cp}=T_{cn}$, or ignored the pressure dependence of $T_{c\alpha}$, or neglected one of the nucleon components.

\textbf{Purpose:} Characterize superfluid-superconducting vortex states at arbitrary pressures with $T_{cp}\neq T_{cn}$, assuming both proton and neutron mean-fields are formed by spin-0 Cooper pairs.

\textbf{Method:} The existing mean-field theory is extended to account for $T_{cp}\neq T_{cn}$. The pressure dependence of the pairing gap energy $\Delta_{\alpha0}$ is quantitatively established on the basis of the effective chiral field theory. To link $T_{c\alpha}$ with $\Delta_{\alpha0}$, I use the weak-coupling result $T_{c\alpha}\approx0.57\Delta_{\alpha0}$. A quadratic scaled-temperature ($T/T_{cp}$) dependence of the thermodynamic magnetic field is postulated in analogy with pure superconductors. The $T/T_{c\alpha}$-dependence of the gap $\Delta_{\alpha T}$ is inferred from the many-body approximations for the pure neutron matter.

\textbf{Results:} An empirical $T/T_{c\alpha}$-dependence for the mean-field is constructed to account for the interplay between the condensation and the magnetic energies.
The superfluid entrainment is found to increase the size of the vortex core and to decrease the effective magnetic penetration depth.
The size of the neutron vortex core is found to be larger than the magnetic penetration depth in the outer core.

\textbf{Conclusions:} The usual approximation of infinitely thin vortex line (the London's approximation) for the neutron vortex is found to be irrelevant in the entire outer core and for the proton vortex is found to be limited to vicinity of the crust-core transition.
The developed mean-field theory paves the way to study the vortex microscopic structure, the angular momentum, the magnetization and the vortex-fluxtube interaction energy.
\end{abstract}

\maketitle

\section{Introduction}
Nucleons in the outer core of neutron stars are believed to compose a liquid similar to nuclear matter in the atomic nucleus.
Observations of neutron stars suggest that the liquid is in the low-$T$ state forming a quantum field with off-diagonal long-range order.
The temperature distribution in the outer core is uniform due to a very high thermal conductivity and thus the matter is isothermal with some value of temperature $T$.
Dynamics of $T$ is slow enough and the thermodynamic equilibrium is established at every instant of time.
Based on microscopic pairing calculations \cite{BPP1969,DrischlerEtal2017,LimHolt2021}, it is expected that the quantum field of nucleons has a mean value which may be treated as a macroscopic variable with characteristic superfluid and superconductor properties, thus supporting quantum vortex states similar to superfluid vortex in Helium and superconducting flux tubes in metals \cite{Anderson1966,Varoquaux2015}.
However, the vortex state in a mixture of superfluid neutrons and superconducting protons has no direct analogue in terrestrial systems and needs to be understood as a unique phenomenon.

The quantum vortex state in the outer core was studied earlier in the literature.
Following the seminal paper \cite{AlparEtal1984} focused on the superfluid entrainment effect in the superfluid-superconducting mixture, many important implications for astrophysical models of pulsars have been investigated.
In studies of the superfluid entrainment effect, a careful distinction between the fluid velocity and momentum is required.
Because of the superfluid entrainment coupling, neutron vortices possess a magnetization as a result of proton current dragged by neutrons and unscreened by the electrons \cite{AlparEtal1984}.

The calculations in \cite{AlparEtal1984} were done in the limit of very low temperatures $T\ll T_{c\alpha}$ neglecting any dependence of the mean fields on $T/T_{c\alpha}$.
For studies of the vortex structure, the Ginzburg-Landau theory was used and according to Sec. II of \cite{AlparEtal1984}, it was a convenience while the calculations and conclusions were not restricted to $T$ being close to either of $T_{c\alpha}$.
It is worth noting though that the order-parameter expansion in the Ginzburg-Landau theory is valid only in vicinity of the critical temperature \cite{Tinkham1996}.
Extension towards the $T=0$ limit can be done in the weak-coupling regime when the Bardeen-Cooper-Schrieffer theory is justified \cite{Tinkham1996}.
However, a continuous connection between the superfluid hydrodynamics and the Ginzburg-Landau theory was missing in the context of superfluid-superconducting nuclear matter.

A similar problem exists in the context of the inner crust.
The vortex structure in the pure neutron matter was addressed in \cite{ElgaroyBlasio2001} when the proton component was neglected.
However, protons in the lower-dimensional, non-spherical, nuclei are expected to be superconducting, thus making the superfluid entrainment effect relevant to the pasta phases \cite{Kobyakov2018,KobyakovPethick2018,PethickZhangKobyakov2020,PethickZhang2021}.
Recently, the proton vortex structure in the inner crust was studied in \cite{PethickZhang2021} when the limit of $T=0$ was considered neglecting the pressure dependence of $T_{c\alpha}$ in the framework of superfluid hydrodynamics.

In the context of the outer core, the mean-field theory developed in \cite{Kobyakov2020} includes the superfluid entrainment and is based on the Ginzburg-Landau expansion in powers of the physical degrees of freedom relevant to the cool nuclear matter in the outer core, the superfluid densities and the superfluid phases generated by the order parameters.
It is designed for temperatures below the smallest of $T_{c\alpha}$ and is valid in the mesoscopic spatial region and for a limited time span.
The mesoscopic extent of the spatial region is defined by the condition $T_{cp}=T_{cn}$.
The limited time span determines the duration when the condition $T(t)\lesssim T_{cp}=T_{cn}$ is fulfilled, where $t$ is the local time.
An interesting alternative approach has been developed in \cite{HaberSchmitt2017} based on power counting of the order parameters.
The superfluid entrainment does not appear in \cite{HaberSchmitt2017} and therefore in the given form it is not directly designed for studies of the quantum vortex states.

If temperature in the outer core is $T\sim 10^8$ K, the assumption $T\ll T_{c\alpha}$ is expected to be the case for the maximum values of $T_{c\alpha}$ in the outer core.
However, the maximum values are separated in the spatial domain of the outer core.
Furthermore, a significant pressure dependence of both $T_{cp}$ and $T_{cn}$ is predicted by modern pairing calculations \cite{DrischlerEtal2017,LimHolt2021}.
There are two consequences of this fundamental property.
First, the condition $T\ll T_{c\alpha}$ fails to universally hold in the outer core, because $T_{cp}$ drops below $T=10^8$ K near the bottom of the outer core, approximately in the region where $T_{cn}$ reaches its maximum.
Second, the assumption $T_{cp}=T_{cn}$ is not satisfied everywhere in the outer core except for a mesoscopic region, a thin layer within the stellar matter, that exists at a particular value of pressure, while the quantum vortex state is expected to extend over a macroscopic region in the outer core.
Therefore, it is of general interest to study the superfluid-superconducting mixture in case when the assumptions $T_{cp}=T_{cn}$ and $T\ll T_{c\alpha}$ are relaxed.

To simplify the calculations and to account for the basic physical features of the superfluid-superconducting mixture in the outer core I will assume that both neutron and proton superfluids are formed by Cooper pairs with total spin zero, corresponding to S-wave pairing.
This approximation in often used in the mean-field theories of the superfluid-superconducting outer core \cite{AlparEtal1984,Link2003,GlampedakisEtal2011,Kobyakov2020,HaberSchmitt2017,WoodGraber2022,ShuklaEtal2024,GranadosEtal2025}.
However, to correctly describe the vortex structure in the outer core and its magnetic and rotational properties, it is important to account for the fact that neutrons are expected to form Cooper pairs with spin-1 \cite{DrischlerEtal2017,MarmoriniYasuiNitta2024}.
It is interesting that spin-1 neutron pairing is also expected in dilute region of dripped neutrons in the inner crust in the magnetic field of suitable strength \cite{TajimaEtal2023}.
The mean-field theory of neutron superfluid with spin-1 pairs has been studied in several works \cite{YasuiEtal2019,MizushimaEtal2021,KobayashiNitta2022,KobayashiNitta2023} neglecting the proton component, with a recent review found in \cite{AdhikariEtal2026}.
The mean-field theory of a mixture of S-wave superconducting protons and P-wave superfluid neutrons applicable in the case $T_{cp}=T_{cn}$ was studied in \cite{HattoriSekizawa2025}.
A future development of a more realistic physical picture of the outer core will require to combine the thermodynamic approach which is the subject of this work, with the mean-field description of proton singlet superconductivity and neutron triplet superfluidity.

The mean-field description for the purpose of this paper is based on the interplay of the condensation and the magnetic energies, while the total density in both nucleon components remains fixed.
The depletion of the superfluid density in the vortex core is not governed by gradients of the nucleon total number densities in agreement with the basic physical picture established in \cite{AlparEtal1984,BlasioElgaroy1999} and with analogous physics of superconductors \cite{Tinkham1996}.
This in contrast to the low-energy modes considered in \cite{KobyakovPethick2017,KobyakovEtAl2017} which are governed by the interplay between the pressure and the matter currents.
Studies of the pure neutron matter have shown that the superfluid vortex core should be considered as filled with normal nucleons \cite{BlasioElgaroy1999} with a small depletion of the total particle density \cite{YuBulgac2003}.
In this respect, the vortex structure in dense Fermi systems such as superfluid nuclear matter is very different from the vortex structure in dense Bose systems such as the superfluid $^4$He \cite{TangEtal2023} in which the vortex core exhibits a complete depletion of the total particle density.

In this paper, a finite-$T$ extension of the quantum vortex state for the case of singlet superconductor mixed with singlet superfluid is constructed on the basis of the mean-field theory developed in \cite{Kobyakov2020}.
I will follow the empirical approach on the basis of thermodynamic arguments and will allow for an arbitrary $T/T_{c\alpha}$-dependence of the superfluid gap
\begin{equation}\label{DeltaT}
  \Delta_{\alpha T}=f_{\alpha T}\Delta_{\alpha 0},
\end{equation}
where $\Delta_{\alpha0}$ are the gap energies calculated at $T=0$ and the $T/T_{c\alpha}$-dependent functions $f_{\alpha T}$ obey the general conditions that $f_{\alpha T}|_{T/T_{c\alpha}=1}=0$ and $f_{\alpha 0}\equiv f_{\alpha T}|_{T/T_{c\alpha}=0}=1$.
For $T/T_{c\alpha}$-dependence of the thermodynamic critical magnetic field, I will postulate the following function motivated by analogy with superconductivity of terrestrial materials:
\begin{equation}\label{HcaT}
  H_{cT}=H_{c0}(1-\tau_{p}^2),
\end{equation}
where $H_{c0}$ is the thermodynamic critical magnetic field at $T=0$ and for convenience, I use the scaled temperature $\tau_\alpha$ according to the definition:
\begin{equation}\label{def_tau}
\tau_\alpha=\frac{T}{T_{c\alpha}}.
\end{equation}
The intensity $H_{cT}$ is defined in the usual way as the value associated with the magnetic energy equal to the difference of the total energy of superconductor in zero magnetic field and the total energy when the superconductivity is completely suppressed by the magnetic field.
The theory approximates the superfluid nuclear matter in the outer core of neutron stats as long as the peculiar effects of triplet neutron superfluidity are not crucial.

The paper is organized as following. Section II is devoted to description of the total energy of the system with emphasis on the superfluid part of the energy. The first law of thermodynamics is used to provide physical meaning of the variables used in the theory. Comparison with alternative formulations of superfluid thermodynamics is provided.
Section III presents the new mean-field equations of motion, which account for $\tau_\alpha$-dependence and the spatial variation of the mean field due to the interplay of the condensation energy, rotation and the magnetic field.
In Sec. IV I derive a concrete form of the coefficients in the mean-field equations on the basis of the assumptions discussed in Eqs. (\ref{DeltaT})-(\ref{def_tau}).
Section V contains a quantitative description of the vortex state.
Summary and conclusions are provided in Sec. VI.

\section{Superfluid thermodynamics}
In order to study characteristic superfluid effects stemming from the nuclear interactions, in particular the superfluid entrainment coupling, I will focus on $T$ below the smallest critical temperature, $T<\min\{T_{cp},T_{cn}\}$.
Quite generally, the total energy
\begin{equation}\label{def_Etot}
E=E^{\mathrm{n}} + E^{\mathrm{s}}
\end{equation}
of nuclear matter in the outer core of neutron stars is a sum of energies of condensation plus the superfluid kinetic energies $E^{\mathrm{s}}$ of the superfluid-superconducting mixture including the magnetic field and the remaining energy due to the underlying Fermi system $E^{\mathrm{n}}$.
In the outer core, the condition $E^{\mathrm{n}} \gg E^{\mathrm{s}}$ always holds.

The total energy $E$ is a function of the outer core total volume $V$, the entropy density $S$, the total particle number densities $n_\alpha$, the normal momenta $\mathbf{P}_\alpha^{\mathrm{n}}$ per particle, the magnetic moment $\mathbf{M}$, the angular momentum $\mathbf{L}$, the order parameters $\psi_{\alpha T}=|\psi_{\alpha T}|\exp(\mathrm{i}\phi_\alpha)$ and its complex conjugate $\psi_{\alpha T}^{*}=|\psi_{\alpha T}|\exp(-\mathrm{i}\phi_\alpha)$ and the magnetic vector potential $\mathbf{A}$:
\begin{equation}\label{def_E}
  E = E(V,S,n_\alpha,\mathbf{P}_\alpha^{\mathrm{n}},\mathbf{M},\mathbf{L},|\psi_{\alpha T}|,\phi_\alpha,\mathbf{A})
\end{equation}
The electron variables are absorbed in the proton labels unless otherwise stated explicitly.
This is convenient, because only stationary states are considered in this paper.
In the stationary case, the electron neutrality and the screening condition \cite{Kobyakov2023} are fulfilled.

The superfluid mean field is a fluctuating quantity and the superfluid density is a concept rather than a rigorously defined number, because the entire ensemble of particles shares both the normal and the superfluid properties \cite{Varoquaux2015}.
In the context of hydrodynamics, the separation of fluid into a superfluid density and the normal density reflects the quantum-mechanical ability of fluid to flow without friction and without a transfer of the entropy \cite{Varoquaux2015}.
This concept is crucial for understanding the fundamental experiments with the superfluids such as the outflow of the liquid formed by $^4$He isotopes (Helium-II) at $T$ below the lambda-point through very narrow capillaries driven by the gravitational force, the motion of Helium-II driven by $T$-gradients, the so-called fountain effect, the Josephson effect for the superfluid phase gradients in superfluids and superconductors and an irrotational character of superfluid flows of Helium-II for slow enough rotation as long as the quantum vortex is not formed.
In the context of superconductivity, the superfluid density measures the singular part located at zero frequency of the oscillator strength in the sum rule applied to the complex electrical conductivity \cite{Tinkham1996}.

A fundamental relation between Helium-II, superconductors and superfluid nuclear matter including pure neutron matter can be seen from the fact that these superfluid systems are described by the same set of equations for the canonically conjugate variables, the superfluid density and the phase, given explicitly below in Eqs. (\ref{equationsSFimp1})-(\ref{equationsSFimp2}).
Importantly, the superfluid phenomenon is not equivalent to the ideal fluid concept in hydrodynamics \cite{Varoquaux2015}.
The reason is simply that the basic degree of freedom in the superfluid is neither the velocity, nor the momentum, but rather it is the superfluid phase of the order parameter.
The superfluid phase is essentially a scalar object and exists in a space of distinct topology as compared to the velocity or momentum vector fields.

The nucleon superfluid mean fields are described by the complex-valued $\tau_\alpha$-dependent wave functions
\begin{equation}\label{def_psia}
  \psi_{\alpha T}(\mathbf{r})=|\psi_{\alpha T}(\mathbf{r})|e^{\mathrm{i}\phi_\alpha(\mathbf{r})}.
\end{equation}
It is convenient to scale $\psi_{\alpha T}(\mathbf{r})$ by their bulk values $n_{\alpha T}\equiv|\psi_{\alpha T}(\infty)|^2$:
\begin{equation}\label{def_nsa}
  \psi_{\alpha T}(\mathbf{r})=\sqrt{n_{\alpha T}}\tilde{\psi}_{\alpha T}(\mathbf{r}),
\end{equation}
where $\tilde{\psi}_{\alpha T}(\infty)=1$ is the scaled value in the bulk and $n_{\alpha T}$ is interpreted as the number density of the Cooper pairs at temperature $T$ in the bulk.
The specific superfluid momentum is defined as
\begin{equation}\label{defPalpha}
  \mathbf{P}_n=\hbar\nabla\phi_n,\quad\mathbf{P}_p=\hbar\nabla\phi_p-\frac{e}{c}\mathbf{A},
\end{equation}
where
\begin{equation}\label{def_phia}
\phi_\alpha=(1/2\mathrm{i})\log(\psi_\alpha/\psi_\alpha^*)
\end{equation}
is the superfluid phase.
The notation in Eqs. (\ref{def_psia})-(\ref{def_phia}) reflects that in the mean-field theory, the $\tau_\alpha$-dependence is inherent to the superfluid densities only, while the superfluid phases are $\tau_\alpha$-independent.

The first law of thermodynamics can be stated in the following form:
\begin{eqnarray}
&& \label{def_dE} dE=\int\,dV \quad TdS-PdV+\mu_\alpha^{\mathrm{tot}}dn_\alpha \\
&& \nonumber  + dA_{\mathrm{lin}}+dA_{\mathrm{ang}}+dA_{\mathrm{mag}} \\
&& \nonumber  + \hbar(\partial_t n_n)d\phi_n - \hbar(\partial_t \phi_n)dn_n \\
&& \nonumber  + (\partial_t n_p)d(\hbar\phi_p-\frac{e}{c}\mathbf{A}) - [\partial_t (\hbar\phi_p-\frac{e}{c}\mathbf{A})]dn_p,
\end{eqnarray}
where $P$ is the pressure, $\mu_\alpha^{\mathrm{tot}}$ are the chemical potentials including the entropy, the pressure and the kinetic energy.

The identifications following from Eq. (\ref{def_dE}),
\begin{eqnarray}
&& \label{equationsSFimp1} \hbar\partial_t\phi_n=-\frac{\partial E}{\partial n_n},\quad \partial_t n_n=\frac{\partial E}{\partial \hbar\phi_n}, \\
&& \label{equationsSFimp2}   \partial_t (\hbar\phi_p-\frac{e}{c}\mathbf{A})=-\frac{\partial E}{\partial n_p},\quad \hbar\partial_t n_p=\frac{\partial E}{\partial (\hbar\phi_p-\frac{e}{c}\mathbf{A})},
\end{eqnarray}
establish an implicit form of the Josephson evolution equation for the superfluid phases and the continuity equation for the superfluid densities.
This form is identical to description of superfluidity and superconductivity in terrestrial samples \cite{Anderson1966,Varoquaux2015}.
The explicit form of Eqs. (\ref{equationsSFimp1}) and (\ref{equationsSFimp2}) for nuclear matter including the superfluid entrainment effect has been discussed earlier in \cite{KobyakovPethick2017} in electrically neutral case and in \cite{Kobyakov2018} in case when the proton electric charge is important.

The contribution from changes in linear momentum can be written as
\begin{equation}\label{dALinMom}
dA_{\mathrm{lin}}=dt\,\sum_{\alpha=n,p}\int\,dV\;\mathbf{j}_{\alpha}^{\mathrm{n}}\cdot\partial_t\mathbf{P}_{\alpha}^{\mathrm{n}}.
\end{equation}
Identification of $dt\,\mathbf{j}_{\alpha}^{\mathrm{n}}$ with the spatial displacement per particle and of $\partial_t\mathbf{P}_{\alpha}^{\mathrm{n}}$ with the corresponding force is consistent with the definition of mechanical work.

The work done by changing the angular momentum is
\begin{equation}\label{dAang}
dA_{\mathrm{ang}} = \pmb{\Omega} \cdot d\mathbf{L},
\end{equation}
where
\begin{equation}\label{def_L}
  \mathbf{L}=\sum_{\alpha=n,p}\int\,dV\;\mathbf{r}\times(\mathbf{j}_\alpha^{\mathrm{n}}+n_\alpha\mathbf{P}_\alpha)
\end{equation}
with $r$ being the radius vector in the cylindrical coordinates with the vertical axis being the axis of rotation.

Finally, I will consider the magnetic energy, which can change $E$ only by the external (free) currents.
The free currents $\mathbf{J}_{\mathrm{free}}$ are not bound to the (otherwise isolated) outer core and are controlled by the observer.
These currents generate a magnetic intensity $\mathbf{H}$ by virtue of the Maxwell equation
\begin{equation}\label{def_H}
\nabla\times\mathbf{H}=\frac{4\pi}{c}\mathbf{J}_{\mathrm{free}}.
\end{equation}
The magnetic intensity is the thermodynamic conjugate variable to the magnetic induction,
\begin{equation}\label{dAmag}
dA_{\mathrm{mag}}=\int\,dV\;\frac{1}{4\pi}\mathbf{H}\cdot d\mathbf{B}.
\end{equation}
The magnetic induction $\mathbf{B}$ is defined as
\begin{equation}\label{def_A}
  \nabla\times\mathbf{A}=\mathbf{B},\quad\nabla\cdot\mathbf{A}=0,
\end{equation}
where the gauge for the vector potential has been chosen explicitly.
A confusion with the work variable $A$ should not arise.
To reveal the physical meaning of the variables $\mathbf{B}$ and $\mathbf{H}$ I use the obvious relation:
\begin{equation}\label{dAmag2}
dA_{\mathrm{mag}}=dt\,\int\,dV\;\frac{1}{4\pi}\mathbf{H}\cdot \partial_t{B}.
\end{equation}
Using the Maxwell equation for the magnetic flux, $\partial_t{B}=-c\nabla\times \mathbf{E}$, where $\mathbf{E}$ is the electric field, and integrating by parts, I find
\begin{equation}\label{dAmag3}
dA_{\mathrm{mag}}=dt\,\int\,dV\;\frac{1}{4\pi}\mathbf{H}\cdot c\nabla\times \mathbf{E}=-dt\,\int\,dV\;\mathbf{J}_{\mathrm{free}}\cdot\mathbf{E},
\end{equation}
where the integration volume includes the system and the observer.
Equation (\ref{dAmag3}) expresses that the system energy $E$ is changed by work of the free currents, fixing the physical meaning of the quantities $\mathbf{H}$ and $\mathbf{B}$.

In terms of superfluid hydrodynamics, the governing equations, which follow from Eq. (\ref{def_dE}) have been studied in \cite{KobyakovPethick2017}.
An alternative form of the first law of thermodynamics was introduced earlier in Eq. (28) of \cite{Mendell1991} in Eq. (69) of \cite{AnderssonComer2006}.
Equation (\ref{def_dE}) of this paper can be rendered to a form similar to Eq. (28) of \cite{Mendell1991} by a partial integration of the term with $\partial n_\alpha$ in Eq. (\ref{def_dE}), where $\partial n_\alpha$ is expressed via the continuity equation and then using the definition of the superfluid momenta in Eq. (\ref{defPalpha}).
The equations of this paper are designed to be used for studies the microscopic vortex structure at arbitrary $T$ in contrast to the equations of \cite{Mendell1991}, when the aim was to study averaged dynamics of a large array of vortices at fixed $T$ and in the limit of $T=0$.

Some differences of the superfluid hydrodynamics on the basis of Eq. (\ref{def_dE}) and that on the basis of Eq. (69) of \cite{AnderssonComer2006} were discussed in \cite{KobyakovPethick2017}, see also \cite{Andersson2021}.
It is interesting to notice that in Eq. (69) of \cite{AnderssonComer2006} the square of the difference of velocities of the two superfluids in Eq. (69) of \cite{AnderssonComer2006} is treated as one of the thermodynamic variables, whereas in this paper the difference of the velocities is determined by the difference of the superfluid phases together with the concrete functional dependence of the superfluid energy $E^{\mathrm{s}}$ on the order parameters $\psi_{\alpha T}$ and $\psi_{\alpha T}^{*}$.

\section{Mean-field equations of motion}
I will consider cases when the total particle densities are fixed, which means there are no dynamic gradients of pressure, in contrast with the cases studied in \cite{KobyakovPethick2017,KobyakovEtAl2017}.
In this paper, the magnetic energy is in interplay with the condensation energy, while the electrons are at rest or uniformly rotating.
In this cases, the stationary states of the superfluid-superconducting mixture are described by the equations
\begin{equation}\label{equationsSF}
  0=\frac{\partial E}{\partial \psi_\alpha},\quad 0=\frac{\partial E}{\partial \psi_\alpha^{*}}.
\end{equation}
At $T$ close to the critical temperature and with the condition $T_{cp}=T_{cn}$, the explicit form of these equations including the entrainment effect was derived in Eqs. (40)-(42) of \cite{Kobyakov2020}.
Additionally, the scalar coupling between the superfluid densities was introduced and the higher-order derivative couplings on the basis of the power counting of the superfluid degrees of freedom were included in \cite{Kobyakov2020}.
Magnitudes of those additional terms were described by the factors $b_{np}$, $\mu_{np}$ and $\lambda_{np}$ and are expected to be small or close to zero depending of details of the nucleon interactions.
Here, I will assume $0=b_{np}=\mu_{np}=\lambda_{np}$.

In order to construct a $\tau_\alpha$-dependent extension of Eqs. (32)-(35) of \cite{Kobyakov2020}, I will account for the $\tau_\alpha$-dependence of the coefficients and the order parameters.
For this purpose, it is convenient to include the subscript $T$ into the notation, which leads to the following changes: $a_\alpha\rightarrow a_{\alpha T}$, $b_\alpha\rightarrow b_{\alpha T}$, $\psi_\alpha\rightarrow \psi_{\alpha T}$.

The superfluid energy functional for a stationary state in presence of the magnetic vector potential $\mathbf{A}$ is a generalization of Eqs. (32)-(35) of \cite{Kobyakov2020}:
\begin{eqnarray}
&& \label{Emeanfield}  E^{\mathrm{s}}=\int\,dV\quad \frac{1}{8\pi}(\nabla \times\mathbf{A})^2 \\
&& \nonumber  + \frac{\hbar^2}{2m}|(\nabla-\mathrm{i}\frac{e}{\hbar c}\mathbf{A})\psi_{pT}|^2 + \frac{\hbar^2}{2m}|\nabla\psi_{nT}|^2 \\
&& \nonumber  + a_{pT}|\psi_{pT}|^2 + \frac{1}{2}b_{pT}|\psi_{pT}|^4 + a_{nT}|\psi_{nT}|^2 + \frac{1}{2}b_{nT}|\psi_{nT}|^4 \\
&& \nonumber  - \frac{\hbar^2}{2m}n_{npT}(\mathbf{P}_p-\mathbf{P}_n)^2,
\end{eqnarray}
where $m$ and $e$ are the mass and the electric charge of the Cooper pair and $n_{npT}$ is the momentum-momentum coupling, the superfluid entrainment, which is directly proportional to the product of the superfluid densities:
\begin{equation}\label{def_nnpT}
  n_{npT}=\frac{c_{np}}{n_0}|\psi_{p T}(\mathbf{r})|^2|\psi_{n T}(\mathbf{r})|^2.
\end{equation}
The quantity $c_{np}$ is a dimensionless factor of the order of unity.
In the limit of $T=0$, the superfluid densities are equal to the uniform nucleon densities, which renders Eq. (\ref{def_nnpT}) to the same form as was used in \cite{KobyakovEtAl2017}.
Therefore, from Eqs. (43) and (50) of \cite{KobyakovEtAl2017} one finds
\begin{equation}\label{cno_range}
-0.5\lesssim c_{np}\lesssim-0.245.
\end{equation}

Introducing the $T$-dependent coefficients and wave functions into Eqs. (40)-(42) of \cite{Kobyakov2020}, I immediately obtain the following equations of motion:
\begin{eqnarray}
&& \label{psip} 0 = -\frac{\hbar^2}{2m}\left(\nabla-\mathrm{i}\frac{e}{\hbar c}\mathbf{A}\right)^2\psi_{pT} + (a_{pT}+b_{pT}|\psi_{pT}|^2)\psi_{pT} \\
&& \nonumber + \mathrm{i}\frac{\hbar}{2m|\psi_{pT}|^2}\nabla\left[n_{npT}(\mathbf{P}_p-\mathbf{P}_n)\right]\psi_{pT} - \frac{1}{2m}\frac{\partial n_{npT}}{\partial |\psi_{pT}|^2}\left(\mathbf{P}_p-\mathbf{P}_n\right)^2, \\
&& \label{psin} 0 = -\frac{\hbar^2}{2m}\nabla^2\psi_{nT} + (a_{nT}+b_{nT}|\psi_{nT}|^2)\psi_{nT} \\
&& \nonumber + \mathrm{i}\frac{\hbar}{2m|\psi_{nT}|^2}\nabla\left[n_{npT}(\mathbf{P}_n-\mathbf{P}_p)\right]\psi_{nT} - \frac{1}{2m}\frac{\partial n_{npT}}{\partial |\psi_{nT}|^2}\left(\mathbf{P}_p-\mathbf{P}_n\right)^2, \\
&& \label{Aeq} \nabla\times\nabla\times\mathbf{A} = \frac{4\pi}{c}\mathbf{J},
\end{eqnarray}
where
\begin{equation}\label{def_J}
\mathbf{J}=e\left[|\psi_{pT}|^2\frac{\mathbf{P}_p}{m}+n_{npT}\frac{\mathbf{P}_n-\mathbf{P}_p}{m}+n_p^{\mathrm{n}}\frac{\mathbf{P}_p^{\mathrm{n}}}{m_p}-n_e\mathbf{v}_e\right]
\end{equation}
is the total electric current including the superconducting proton, the normal proton and the electron contributions with $m_p$ being the proton mass and $\mathbf{v}_e$ being the electron fluid velocity.

It follows from Eqs. (\ref{psip})-(\ref{psin}) that the bulk values are given by the following expressions:
\begin{equation}\label{bulks}
  |\psi_{\alpha T}(\infty)|^2=n_{\alpha T}=-\frac{a_{\alpha T}}{b_{\alpha T}}.
\end{equation}
In the absence of rotation of the electrons, $\mathbf{P}_\alpha(\infty)=0$ and $\mathbf{A}(\infty)=0$.
If the electrons rotate with angular velocity $\pmb \Omega$, then the condition of the absence of the electric current in the bulk, $\mathbf{J}(\infty)=0$, is equivalent to
\begin{equation}\label{ALondon}
\mathbf{A}(\infty)=\frac{mc}{e}\frac{n_p^{\mathrm{n}}(T)-n_e}{n_{pT}-n_{npT}}\pmb \Omega\times \mathbf{r},
\end{equation}
where the London magnetic field \cite{AlparEtal1984} is equal to
\begin{equation}\label{def_BL}
\mathbf{B}_L=\nabla\times\mathbf{A}(\infty),
\end{equation}
the electron number density is $n_e$ and
\begin{equation}\label{npnorm}
n_p^{\mathrm{n}}(T)=2n_{p0}-2n_{pT}
\end{equation}
is the normal proton number density.
The number densities satisfy the electric neutrality condition that $n_e=n_p^{\mathrm{n}}(T_{cp})=2n_{p0}$.
The screening condition is that the electrons follow the motion of the normal protons, ${\mathbf{P}_p^{\mathrm{n}}}/{m_p}=\mathbf{v}_e$.
Equation (\ref{def_BL}) generalizes the earlier results for the London field \cite{AlparEtal1984,Mendell1991,GlampedakisEtal2011} to the case of arbitrary $T$.

\section{Pressure dependence of the superfluid mean field}
In the outer core, the superfluid mean field is nonuniform.
The reason is not only that the nucleon densities are pressure-dependent.
Importantly, the spatial dependence of the nucleon density in the outer core is not replicated by the spatial dependence of the superfluid density, because the latter is determined by Eq. (\ref{bulks}) and is not equal to the former everywhere in the outer core even at the lowest typical temperature $T\sim10^8$ K.
Given the direct correspondence between the pressure and spatial position in the outer core, it is convenient to focus on the pressure dependence of $T_{c\alpha}$.
In turn, finding the pressure dependence of $T_{c\alpha}$ is equivalent to establishing the dependence of $T_{c\alpha}$ on the total baryon density $n_b$.

The quantity $T_{c\alpha}$ has not been directly calculated for the range of $n_b$ in the outer core.
To circumvent this difficulty, I will utilize the weak-coupling pairing approximation \cite{Tinkham1996} and calculate $T_{c\alpha}$ from the knowledge of the superfluid gap at $T=0$, $\Delta_{\alpha0}\equiv\Delta_{\alpha T}|_{T=0}$:
\begin{equation}\label{TcDeltaa}
T_{c\alpha}\approx0.57\Delta_{\alpha0}.
\end{equation}
Whether Eq. (\ref{TcDeltaa}) is a good approximation for pure neutron matter was studied recently in \cite{DrissiRios2022}.
It was found that due to a mutual cancelling of the strong interaction effects, the ratio $T_{c\alpha}/\Delta_{\alpha0}$ is indeed almost constant close to 0.57.
Another useful result reported in \cite{DrissiRios2022} is the $T/T_{c}$-dependence of the neutron pairing gap energy $\Delta(T/T_{c})$, were $T_{c}$ is the superfluid transition temperature in pure neutron matter.
The $T/T_{c}$-dependence of the superfluid gap is contained a function equal to the ratio $\Delta_{\alpha T}/\Delta_{\alpha0}$ between the superfluid gap at $T$ and its $T=0$ limit.
Since the results of \cite{DrissiRios2022} on pure neutron matter are the only available quantitative study, I will assume that the same form of the function $\Delta_{\alpha T}/\Delta_{\alpha0}$ holds for nuclear matter.

In contrast to $\tau_\alpha$-dependence, the dependence of the gap on the baryon density (and hence on the pressure) has been studied extensively.
I will utilize the data on $\Delta_{\alpha0}$ as function of $n_b$, obtained earlier in the framework of the effective chiral field theory.
The most recent calculations have been reported in \cite{DrischlerEtal2017} for neutron pairing and \cite{LimHolt2021} for proton pairing.
Using the beta-equilibrium proton chemical fraction at given values of the proton Fermi wave number $x(k_{Fp})$, the pairing gaps $\Delta_{\alpha0}(k_{F\alpha})$ reported in \cite{DrischlerEtal2017} and \cite{LimHolt2021} can be represented as functions of the total baryon density $\Delta_{\alpha0}(n_b)$.
Using $k_{Fp}^3 = 3\pi^2xn_b$ and $k_{Fn}^3 = 3\pi^2(1-x)n_b$ where $n_b$ is the total baryon density, and combining the data reported in Fig. 7 of \cite{LimHolt2021} and those in panel at the center column in the third row of Fig. 7 of \cite{DrischlerEtal2017}, it is straightforward to evaluate characteristic numerical values $\Delta_{\alpha0}(n_b)$ and plot them on a common graph.
The theoretical uncertainties were given in \cite{DrischlerEtal2017} and \cite{LimHolt2021} and can be readily included in the calculations in this paper.
For the sake of clarity in this paper, I will not include the uncertainty estimates, but rather will focus on construction of a new phenomenological method to account for the unequal critical temperatures with different pressure dependence.

Figure 1 shows $T_{c\alpha}$ as function of $n_b$ obtained from the data on $\Delta_{\alpha0}$ combined with Eq. (\ref{TcDeltaa}).
The values shown in Fig. 1 will be used for the numerical evaluation in this paper.
It is interesting that the data displayed in Fig. 1 provides noticeable corrections to the usual evaluation of nuclear properties at the nuclear saturation density $n_0$ \cite{AlparEtal1984,Link2003,GlampedakisEtal2011,Kobyakov2023}.
Here I will use the value obtained from the nuclear interaction parametrization Sk$\chi$450 yielding $n_0=0.1561$ fm$^{-3}$.
\begin{figure}
\includegraphics[width=3.5in]{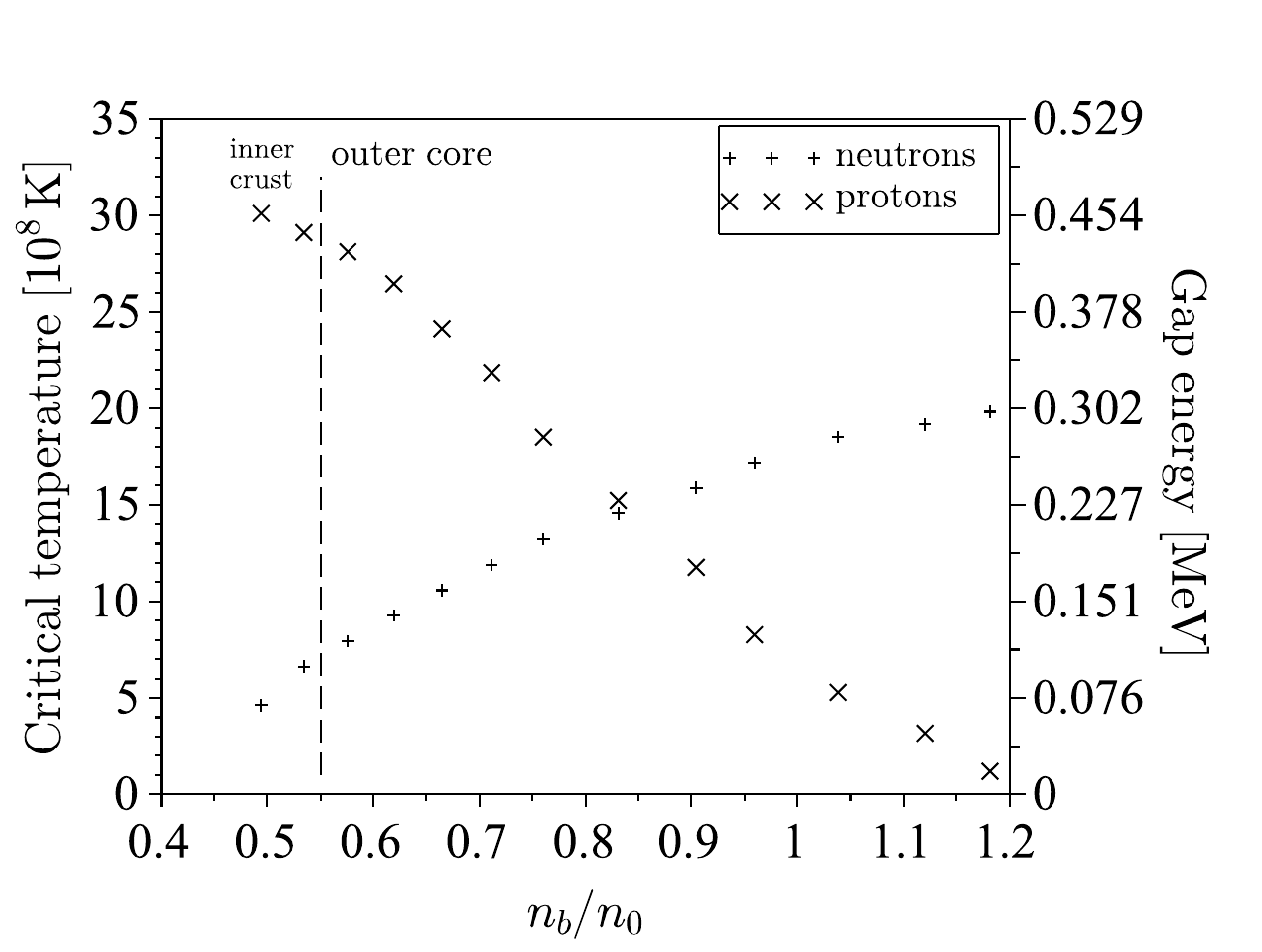}
\caption{Critical temperatures $T_{c\alpha}$ of neutron and proton pairing in the weak pairing approximation inferred from Eq. (\ref{TcDeltaa}) and the data on the pairing gap energy at $T=0$, $\Delta_{\alpha0}$, reported in \cite{DrischlerEtal2017} and \cite{LimHolt2021} on the basis of the effective chiral field theory. The vertical dashed line shows the critical baryon density, below which uniform nuclear matter of the outer core is unstable with respect to density fluctuations and the inner crust region is expected.}
\end{figure}

It remains to establish a link between the functions $\Delta(T)$ and $\Delta_\alpha(n_b)$ with the coefficients $a_{\alpha T}$ and $b_{\alpha T}$ in the mean-field Eqs. (\ref{psip})-(\ref{Aeq}).
In the existing approach to superconductivity in terrestrial materials at arbitrary $T$ below the critical temperature \cite{Tinkham1996}, the $T$-dependence of the mean-field is inferred from the thermodynamic critical magnetic field $H_c(T)$ and the effective magnetostatic penetration depth $\lambda^{\rm eff}(T)$, which are both experimentally accessible.
This results in the following dependence of the superconducting density $n^{\rm s}$ and the mean-field coefficients of the Ginzburg-Landau theory in the weak-coupling approximation $a(T)$ and $b(T)$: $n^{\rm s}={m^{*}c^2}/{8\pi {e^{*}}^2(\lambda^{\rm eff})^2}$, $a(T)=-({2{e^{*}}^2}/{m^{*}c^2})(H_c\lambda^{\rm eff})^2$, $b(T)=({16\pi {e^{*}}^4}/{{m^{*}}^2c^4})H_c^2(\lambda^{\rm eff})^4$,
where $e^{*}$ and $m^{*}$ are the electric charge and the mass of the Cooper pair.
Note that in pure superconducting medium, which is the case for the nucleon superfluids in the outer core, as opposed to dirty medium, the superconducting density $n_{p0}^{\rm s}$ at $T=0$ is equal to half the total number density of the charge carriers (the electrons in superconductors and the protons in the outer core).

For terrestrial superconductors, the empirical approximations have been postulated as $H_c(T)=H_c(0)(1-\tau^2)$ (this motivates my initial assumption in Eq. (\ref{HcaT})), and $\lambda^{\rm eff}(T)=\lambda^{\rm eff}(0)/\sqrt{1-\tau^4}$ \cite{Tinkham1996}.
Unfortunately, this scheme cannot be directly transferred to the mean-field theory of the superfluid-superconducting outer core, because neither the critical magnetic field $H_{cT}$, nor the effective magnetostatic penetration depth $\lambda^{\rm eff}_T$ are accessible experimentally.
Additionally, the superfluid densities $n_{\alpha T}^{\rm s}$ normalized in suitable units to the pairing gaps are expected to have a nontrivial $\tau_\alpha$-dependence.

An empirical construction can be done by using the data on $\Delta(T)$ obtained in \cite{DrissiRios2022}.
Based on calculations for the pure neutron matter \cite{DrissiRios2022}, I will postulate the following phenomenological dependence:
\begin{eqnarray}
&& \label{fTTc} f_{\alpha T} = c_1 \sqrt{({1-\tau_\alpha^2})/({1+\tau_\alpha^2})},\quad(T\lesssim T_{c\alpha}) \\
&& \label{fT0} f_{\alpha T} =  1-\exp(-c_2/\tau_\alpha),\quad(T\gtrsim 0).
\end{eqnarray}
The function in Eq. (\ref{fTTc}) is consistent with the characteristic behavior of all mean-field theories \cite{Tinkham1996}.

The data shown in left panel of Fig. 1 of \cite{DrissiRios2022} corresponds to the weak-coupling (BCS) plus Hartree-Fock model and will be used here to infer the functions $f_\alpha(T)$.
Figure 2 shows the data on $\Delta(T)$ adopted from \cite{DrissiRios2022} with the fits according to Eqs. (\ref{fTTc}) and (\ref{fT0}) with $c_1=1.4$ and $c_2=1.5$.
Although there is a cusp near $n_b/n_0\sim0.64$, I will assume that the approximation by Eq. (\ref{fT0}) at $n_b/n_0\leq0.64$ and by Eq. (\ref{fTTc}) at $n_b/n_0\geq0.64$ are approximately valid.
\begin{figure}
\includegraphics[width=3.5in]{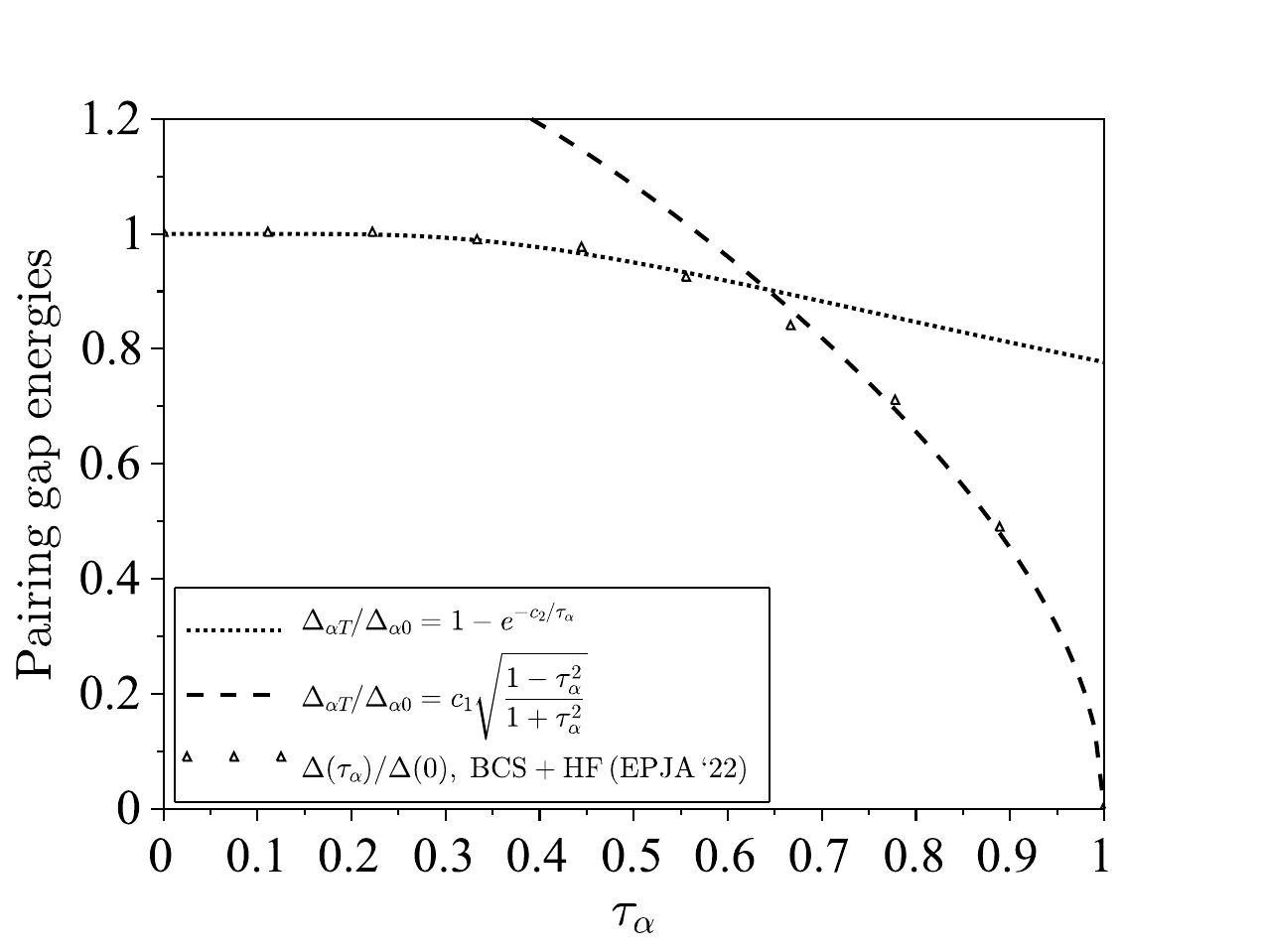}
\caption{Triangles: Dependence of the scaled pairing gap $\Delta(\tau_\alpha)/\Delta(0)$ in pure neutron matter corresponding to the weak-coupling (BCS) plus Hartree-Fock model, where $\tau_\alpha$ corresponds to $T/T_c$, with $T_c=0.57\Delta(0)$ and $\Delta(T)$ adopted from Fig. 1 of \cite{DrissiRios2022} with the EM+3NF force. Dashed and dotted lines show the numerical fit corresponding to Eq. (\ref{fTTc}) with $c_1=1.4$ and Eq. (\ref{fT0}) with $c_2=1.5$.}
\end{figure}

Having postulated the empirical approximations for $\Delta_{\alpha T}$, it remains to postulate approximations for the mean-field quadratic coefficient $a_{\alpha T}$ and the quartic coefficient $b_{\alpha T}$.
For this purpose, I assume that the weak-coupling expression for the coherence length is valid:
\begin{equation}\label{ksiaT}
  \xi_{\alpha T}=\frac{\hbar v_{F\alpha}}{\pi \Delta_{\alpha T}},
\end{equation}
where $v_{F\alpha}=\hbar k_{F\alpha}/m_p$ is the Fermi velocity.
Equations (\ref{DeltaT}) and (\ref{ksiaT}) fix the $\tau_\alpha$-dependence of $\xi_{\alpha T}$.
Therefore, the $\tau_\alpha$-dependence of $a_{n T}$ is as well fixed, because $\xi_{\alpha T}$ is defined by the mean-field equations as
\begin{equation}\label{def_xiaT}
\xi_{\alpha T}^2=-\hbar^2/2ma_{\alpha T}.
\end{equation}
For protons, I will assume that both $a_{p T}$ and $b_{p T}$ are $\tau_\alpha$-dependent.
For neutrons, I will assume that only $a_{n T}$ bears the $\tau_\alpha$-dependence, while $b_{n T}$ is $\tau_\alpha$-independent.

Using Eqs. (\ref{DeltaT}), (\ref{ksiaT}) and (\ref{def_xiaT}) and the thermodynamic definition
$H_{cT}^2=4\pi \frac{a_{p T}^2}{b_{p T}}$,
the coefficient $b_{p T}$ is found.
Finally, a straightforward generalization of the London penetration depth
$\lambda_{T}^2=\frac{mc^2}{4\pi e^2}\frac{b_{p T}}{-a_{p T}}$
can be calculated. Collecting the analytical results, I find
\begin{eqnarray}
&& \label{def_aaT} a_{\alpha T}=-\frac{\pi^2\Delta_{\alpha T}^2}{2mv_{F\alpha}^2}, \\
&& \label{def_bpT} b_{p T}=\frac{\pi^2\Delta_{pT}^4}{2mv_{Fp}^2\Delta_{p0}^2n_{p0}(1-\tau_p^2)^2}, \\
&& \label{def_bnT} b_{n T}=\frac{\pi^2\Delta_{n0}^2}{2mv_{Fn}^2n_{p0}}, \\
&& \label{def_HcT} H_{c T}=\sqrt{\frac{2\pi^3n_{p0}}{m}}\frac{\Delta_{p0}}{v_{Fp}}(1-\tau_p^2), \\
&& \label{def_lambdaT} \lambda_{T}=\sqrt{\frac{mc^2}{4\pi e^2 n_{p0}}}\frac{\Delta_{pT}}{\Delta_{p0}}(1-\tau_p^2)^{-1}.
\end{eqnarray}
As we will see below, $\lambda_{T}$ is a convenient theoretical parameter, which differs from the actual penetration depth $\lambda_{T}^{\mathrm{eff}}$ due to the superfluid entrainment.
Equations (\ref{def_aaT})-(\ref{def_lambdaT}) completely define the mean-field equations of motions Eqs. (\ref{psip})-(\ref{Aeq}) and may be used to study the vortex structure.

\section{Mean-field description of the vortex state}
It is convenient to use the dimensionless variables, where the wave functions are scaled according to Eq. (\ref{def_nsa}), the magnetic vector potential is scaled as $\mathbf{A}=H_{cT}\xi_{nT}\tilde{\mathbf{A}}$ and the coordinate is scaled by the neutron coherence length $\mathbf{r}=\xi_{nT}\tilde{\mathbf{r}}$ and thus $\nabla=\xi_{nT}^{-1}\tilde{\nabla}$.
Notice that in Eqs. (\ref{psip})-(\ref{psin}), the terms proportional to $\mathrm{i}{\hbar}/{2m|\psi_{\alpha T}|^2}$ have been dropped because those are always zero for straight purely proton or purely neutron vortex configurations considered in this paper.

Dropping the tildes from the dimensionless variables $\tilde{\psi}_{\alpha T}$, $\tilde{\mathbf{A}}$ and $\tilde{\mathbf{r}}$, I cast Eqs. (\ref{psip})-(\ref{Aeq}) into the following dimensionless form (for the sake of clarity I forget for a moment about the normal fluid motion, which can be readily incorporated into the equations of motion):
\begin{eqnarray}
&& \label{psipDL} 0 = \left(\nabla-\frac{\mathrm{i}}{\gamma_0}\mathbf{A}\right)^2\psi_{pT} + \gamma_1(1-|\psi_{pT}|^2)\psi_{pT}  \\
&& \nonumber - \gamma_2 |\psi_{nT}|^2\left(\nabla(\phi_p-\phi_n)-\frac{1}{\gamma_0}\mathbf{A}\right)^2, \\
&& \label{psinDL} 0 = \nabla^2\psi_{nT} + (1-|\psi_{nT}|^2)\psi_{nT} \\
&& \nonumber + \gamma_3|\psi_{pT}|^2\left(\nabla(\phi_p-\phi_n)-\frac{1}{\gamma_0}\mathbf{A}\right)^2, \\
&& \label{AeqDL} \nabla\times\nabla\times\mathbf{A} = \frac{\xi_{nT}^2}{\lambda_T^2}|\psi_{pT}|^2 \\
&& \nonumber \times\left[\gamma_0\nabla\phi_p-\mathbf{A}+\gamma_0\gamma_2|\psi_{nT}|^2\left(\nabla(\phi_n-\phi_p)+\frac{1}{\gamma_0}\mathbf{A}\right)\right],
\end{eqnarray}
where
\begin{eqnarray}
&& \label{gamma01} \gamma_0=\frac{\hbar c}{eH_{cT}\xi_{nT}^2}, \quad \gamma_1=\frac{a_{pT}}{a_{nT}} \\
&& \label{gamma23} \gamma_2=\frac{c_{np}n_{nT}}{n_0}, \quad \gamma_3=\frac{c_{np}n_{pT}}{n_0}.
\end{eqnarray}
The characteristic lengths of the vortex state can be readily obtained from Eqs. (\ref{psipDL})-(\ref{gamma23}).
The dimensionless coefficients and the thermodynamic critical magnetic field are evaluated for $T=0$ in Fig. 3.
\begin{figure}
\includegraphics[width=3.5in]{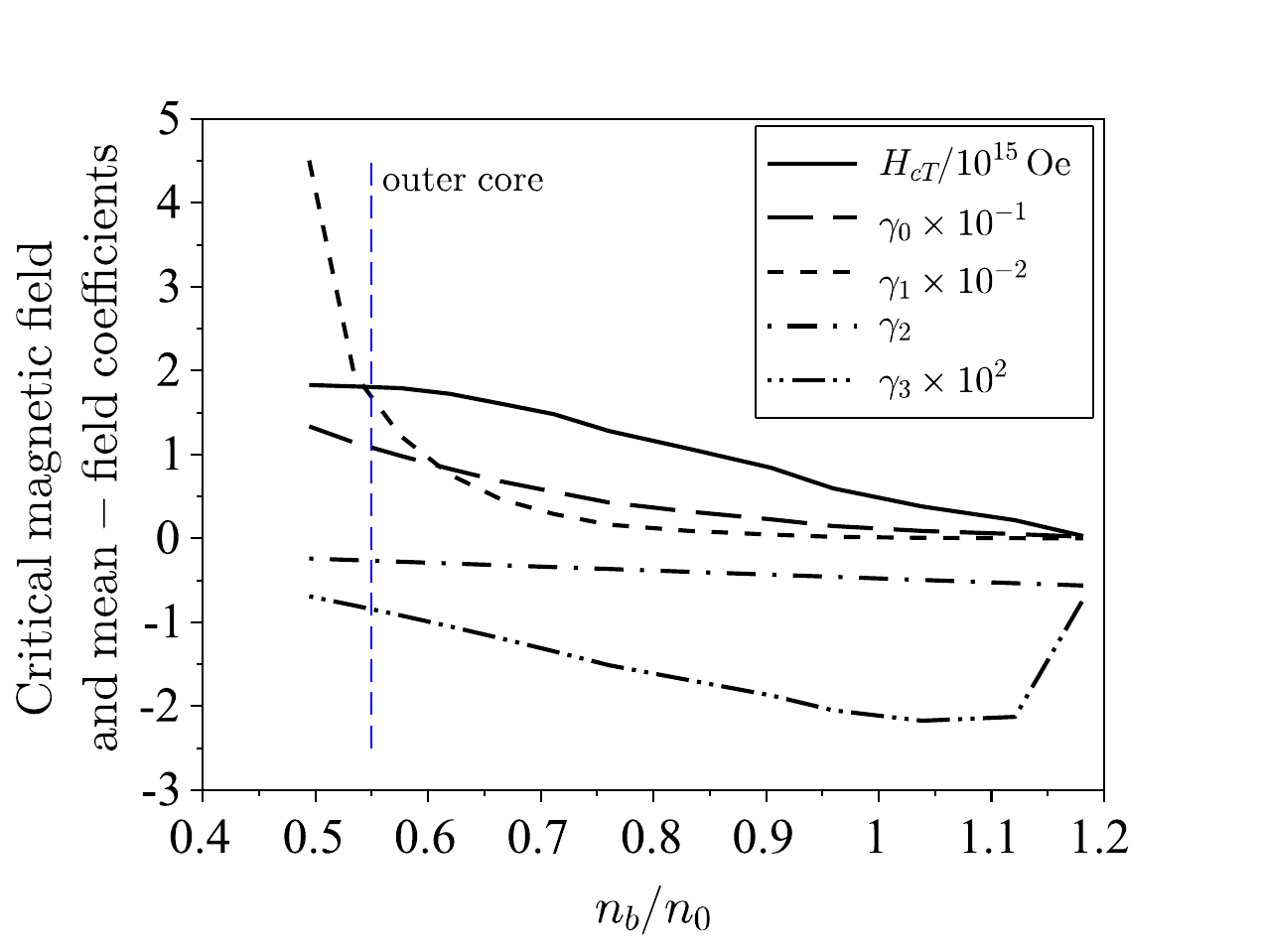}
\caption{The dimensionless coefficients $\gamma_0$, $\gamma_1$, $\gamma_2$, $\gamma_3$, Eqs. (\ref{gamma01})-(\ref{gamma23}), and the thermodynamic critical magnetic field, Eq. (\ref{def_HcT}). Calculations are done with $T=10^8$ K. The data relevant to the outer core are located to the right from the blue dashed line. }
\end{figure}
\subsection{Neutron vortex}
In purely neutron vortex, the wave functions have the form $\psi_{\alpha T}=|\psi_{\alpha T}(r)|\exp(\mathrm{i}\nu_\alpha \varphi)$, where $r$ and $\varphi$ are the radial and angular coordinates in the cylindrical coordinates with the vertical axis coinciding with the vortex axis and $\nu_\alpha$ is the vortex quantum winding number with $\nu_n=\pm1,\pm2,\ldots$ and $\nu_p=0$.
The boundary conditions are $0=\psi_{n T}(0)=A(0)=A(\infty)$ and $1=\psi_{n T}(\infty)=\psi_{p T}(\infty)$.
The laplacian term yields
\begin{equation}\label{laplacian_psin}
\nabla^2\psi_{nT}=\frac{1}{r}\partial_r(r\partial_r\psi_{nT})-\frac{\nu_n^2}{r^2}\psi_{nT}.
\end{equation}
and the entrainment term yields
\begin{equation}\label{entr_psin}
  \gamma_3\left(\nabla(\psi_p-\phi_n)-\frac{1}{\gamma_0}\mathbf{A}\right)^2=\gamma_3\left(\frac{\nu_p-\nu_n}{r}-\frac{1}{\gamma_0}\mathbf{A}\right)^2.
\end{equation}
Near the vortex axis, in vicinity of $r=0$ these two terms and the linear term in Eq. (\ref{psinDL}) dominate and satisfy the equation
\begin{equation}\label{nvortexaxis}
  0=\left(-\frac{\nu_n^2}{r^2}+\gamma_3\frac{\nu_n^2}{r^2}+1\right)\psi_{nT}.
\end{equation}
Therefore, the neutron vortex core is characterized by a length scale $\tilde{r}_{nT}=\nu_n\sqrt{1-\gamma_3}$.
The dimensional form of the neutron vortex core radius is
\begin{equation}\label{rn}
  r_{nT}=\xi_{nT}\nu_n\sqrt{1-\gamma_3}.
\end{equation}
\subsection{Proton vortex}
In purely proton vortex the winding numbers are $\nu_p=\pm1,\pm2,\ldots$ and $\nu_n=0$.
The further consideration is analogous to the neutron vortex case.
The boundary conditions are $0=\psi_{p T}(0)=A(0)=A(\infty)$ and $1=\psi_{p T}(\infty)=\psi_{n T}(\infty)$.
Near $r=0$, Eq. (\ref{psipDL}) reduces to
\begin{equation}\label{pvortexaxis}
  0=(-\frac{\nu_p^2}{r^2}+\gamma_2\frac{\nu_p^2}{r^2}+\gamma_1)\psi_{nT}.
\end{equation}
Using the relation $\xi_{nT}/\sqrt{\gamma_1}=\xi_{pT}$, I find that the proton vortex core radius is
\begin{equation}\label{rp}
  r_{pT}=\xi_{pT}\nu_p\sqrt{1-\gamma_2}.
\end{equation}
\subsection{The magnetic penetration depth}
In order to obtain the asymptotic form of Eq. (\ref{AeqDL}) in the bulk, it is convenient to work with the dimensionless magnetic induction $\nabla\mathbf{A}=\mathbf{B}=\mathbf{e_z}B(r)$.
Application of curl to Eq. (\ref{AeqDL}) and using the boundary conditions in the bulk, $1=\psi_{p T}(\infty)=\psi_{n T}(\infty)$, yields
\begin{equation}\label{Avortexbulk}
  0=r^2\partial_{rr}^2B+r\partial_rB-\frac{\xi_{nT}^2}{\lambda_T^2}(1-\gamma_2)r^2B,
\end{equation}
which is the modified Bessel equation. According to the boundary condition in the bulk, $0=A(\infty)=B(\infty)$, the relevant solution is of the second kind, which is exponentially decaying with the characteristic decay rate $\lambda_T/\xi_{nT}\sqrt{1-\gamma_2}$.
The magnetic penetration depth is therefore
\begin{equation}\label{lambdaeff}
  \lambda_{T}^{\mathrm{eff}}=\frac{\lambda_{T}}{\sqrt{1-\gamma_2}}.
\end{equation}
\subsection{Numerical results}
The characteristic lengths associated with the superfluid vortex state can now be computed with the chosen model for the nuclear equation of state from Eqs. (\ref{ksiaT}), (\ref{def_lambdaT}), (\ref{rn}), (\ref{rp}) and (\ref{lambdaeff}).
Setting $T=10^8$ K and using the baryon-density dependence of the gaps displayed in Fig. 1 and that of $T_{c\alpha}$ using Eqs. (\ref{fTTc}) and (\ref{fT0}), I plot the results for $\xi_{\alpha T}$, $r_{\alpha T}$, $\lambda_{T}$ and $\lambda_{T}^{\mathrm{eff}}$ in Fig. 4.
\begin{figure}
\includegraphics[width=3.5in]{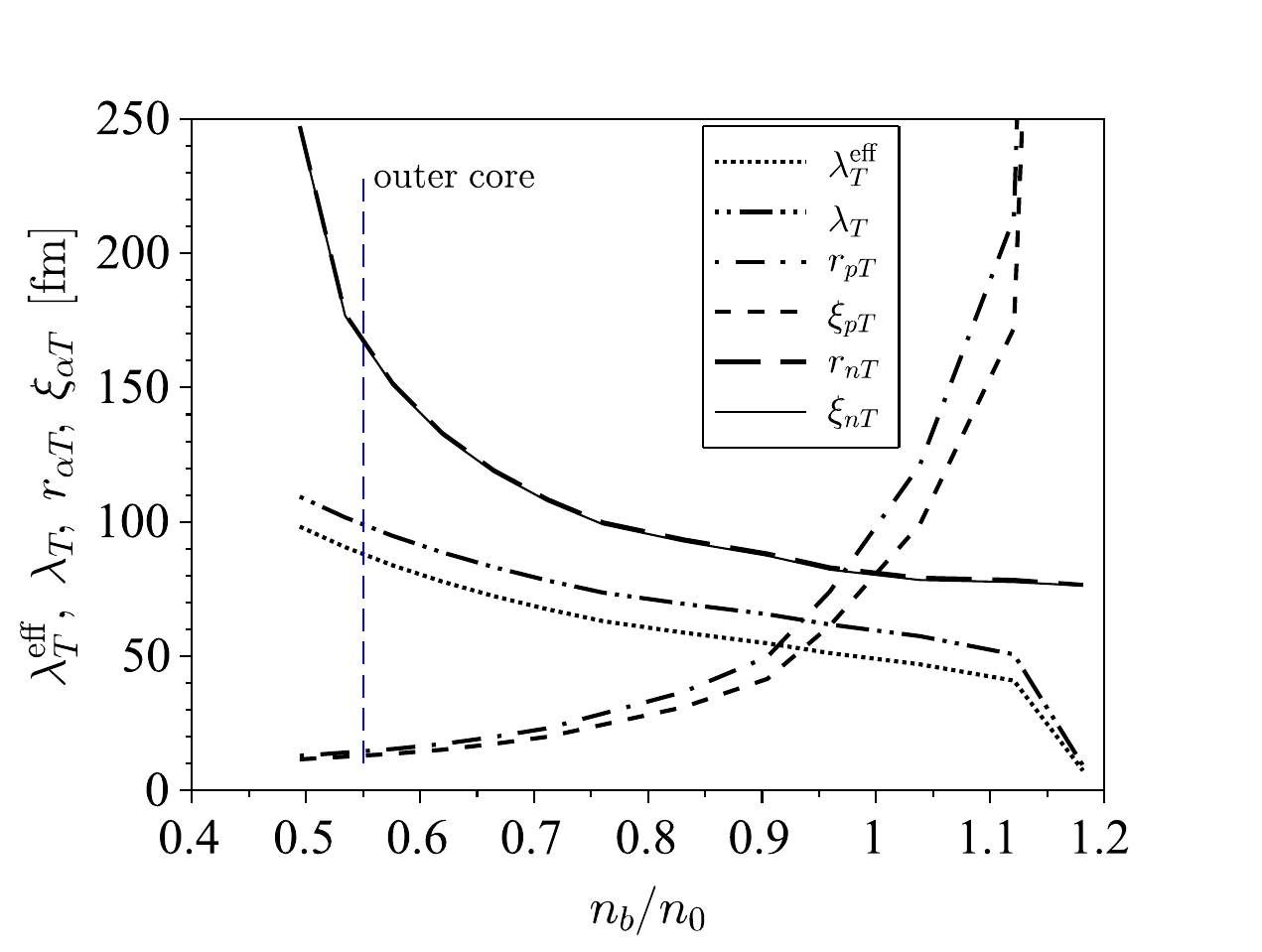}
\caption{Characteristic lengths associated with the superfluid vortex state: the effective magnetostatic penetration depth $\lambda_{T}^{\mathrm{eff}}$, Eq. (\ref{lambdaeff}), the core radius of purely neutron $r_{n T}$, Eq. (\ref{rn}), or purely proton $r_{p T}$, Eq. (\ref{rp}), vortex. For comparison, the coherence lengths of proton superconductor $\xi_{p T}$ and of neutron superfluid $\xi_{n T}$ are shown together with the intrinsic theoretical parameter $\lambda_{T}$, Eqs. (\ref{def_xiaT})-(\ref{def_lambdaT}). Calculations are done with $T=10^8$ K. The data relevant to the outer core are located to the right from the blue dashed line. }
\end{figure}

Figure 4 highlights the difference between the intrinsic theoretical parameters ($\xi_{\alpha T}$ and $\lambda_{T}$) introduced for mathematical convenience and the actual physical quantities ($r_{\alpha T}$ and $\lambda_{T}^{\mathrm{eff}}$), which bear the physical significance and directly characterize the vortex state.
If the system were available for the experimental measurements, the quantities $\lambda_{T}^{\mathrm{eff}}$ and $H_{cT}$ would be observable.
Nevertheless, it is instructive to examine the both sets $\xi_{\alpha T}$, $\lambda_{T}$ and $r_{\alpha T}$, $\lambda_{T}^{\mathrm{eff}}$ on the common figure.
This shows that the quantitative effect of the superfluid entrainment is to decrease the magnetic penetration depth and to increase the core radius of the purely proton vortex. The core radius of the purely neutron vortex is seen to be almost unaffected by the entrainment.
The difference of the modifications of the core radii can be understood as a consequence of a higher density of neutron pairs as compared to proton pairs everywhere in the outer core.

\section{Summary and conclusions}
In this paper, I have introduced a mean-field theory of superfluid-superconducting mixture relevant to the outer core of neutron stars.
The mean field is understood as the complex-valued macroscopic wave function containing the information on the superfluid density $|\psi_{\alpha T}|^2$ and the superfluid phase $\phi_\alpha$.
The superfluid energy functional $E^{\mathrm{s}}$ has been obtained by the expansion of energy in powers of the most relevant superfluid degrees of freedom, $|\psi_{\alpha T}|^2$ and $\nabla\phi_\alpha$, and therefore represents a well-defined mean-field superfluid-superconducting theory.
An outstanding feature of the theory is that it correctly matches the superfluid hydrodynamic theory in the $T=0$ limit in the crucial aspect of the superfluid entrainment effect.
This is achieved by using a Galilean-invariant contribution directly proportional to $|\psi_{p T}|^2$, $|\psi_{n T}|^2$ and $(\nabla\phi_p-\nabla\phi_n)^2$.
The magnetic interaction has been included using the electromagnetic gauge-invariant approach.

Unveiling the $T$-dependence and the $n_b$-dependence of the mean-field energy functional had been the main problem solved in this work.
Thus, a continuous connection between the two limiting cases (the superfluid hydrodynamics valid at zero temperature and the Ginzburg-Landau theory valid near the smallest critical temperature) is established.
The solution is based on an empirical approach in close analogy with the mean-field theories of superconductivity.
The starting point was to assume a concrete $\tau_\alpha$-dependence of two fundamental quantities in the theory: The pairing gaps energy and the thermodynamic magnetic field, Eqs. (\ref{DeltaT}) and (\ref{HcaT}).
Next, using the weak-coupling results (from the Bardeen-Cooper-Schrieffer theory) on the relation between the critical temperature and the gap energy, Eq. (\ref{TcDeltaa}), together with the relation between the coherence length and the gap energy, Eq. (\ref{ksiaT}), the coefficients $a_{pT}$ and $b_{pT}$ of the proton superconductor mean-field have been determined.
For the neutron superfluid mean-field, the scheme was slightly different because the thermodynamic magnetic field is irrelevant.
Because of this, it has been convenient to assume that the quartic coefficient $b_{nT}$ is $\tau_\alpha$-independent.
This scheme has allowed to completely determine the mean-field superfluid energy functional $E^{\mathrm{s}}$ in Eq. (\ref{Emeanfield}) leading to the equations of motion, Eqs. (\ref{psip})-(\ref{Aeq}).

In conclusion, the mean-field theory developed in this paper has allowed to characterize the vortex state in superfluid-superconducting mixture relevant to the outer core on the basis of the data on the pairing gap adopted from the earlier literature and shown in Figs. 1 and 2.
The mean-field equations derived in this paper can be readily used for evaluation of the vortex magnetization and the angular momentum as well as for calculation of the vortex-fluxtube interaction energy.

Figure 3 shows that even at the lowest temperature expected in the outer core, $T=10^8$ K, the coefficients in the mean-field equations differ significantly from constant values.
The spatial gradients of the coefficients are related to variations on the scale of the spatial extent of the outer core and thus are negligibly smaller than the variations of the order parameters and of the magnetic field near the vortex line.
Nevertheless, the inhomogeneity of the mean field that follows by combining the data from Fig. 3 with the stellar pressure profile, is important for quantitative description of the vortex-fluxtube interactions and of the rotational and magnetic properties of the vortex states, which extend throughout the outer core.

The main numerical results of this paper are shown in Figs. 3 and 4.
It is found that the neutron vortex core radius $r_{n T}$ is larger than the effective magnetic penetration depth $\lambda_{T}^{\mathrm{eff}}$ everywhere in the outer core.
This finding reveals new physics of superfluid neutron vortices in the outer core of neutron stars.
For the neutron vortex, evaluation of its structure would require to relax the London's approximation of infinitely thin vortex line used in the earlier literature.
Because of this, the neutron vortex core structure should be studied in terms of the system of coupled Eqs. (\ref{psipDL})-(\ref{AeqDL}), which requires a numerical solution and goes beyond the aims of this work.

For the proton vortex, Fig. 4 confirms the validity near the crust-core transition region of the London's approximation used in the earlier studies.
On the basis of the chosen equation of state, Fig. 4 predicts a transition between type-I and type-II superconductivity in accordance with conclusions of the earlier studies.
The results shown in Fig. 4 reveal that in the region of baryon density $0.8n_0\lesssim n_b$ the London's approximation is not applicable to proton vortex and its structure should be studied numerically by solving Eqs. (\ref{psipDL})-(\ref{AeqDL}).

The theory may be further improved in several directions.
The dependence of the gap on $T$, Eq. (\ref{DeltaT}), should be calculated for the conditions of the outer core.
This would allow to use the most relevant functions $f_{\alpha T}$ instead of using a particular function found in case of pure neutron matter.
Also, it is reasonable to verify the basic empirical assumption of the dependence of the thermodynamic critical magnetic field on $\tau_\alpha$, Eq. (\ref{HcaT}), by direct microscopic calculations.
It should be useful to confirm or disprove by direct microscopic calculations the assumptions regarding the gradient and the amplitude couplings ($0=b_{np}=\mu_{np}=\lambda_{np}$) in the superfluid energy functional $E^{\mathrm{s}}$.
Finally, a crucial step towards a realistic mean-field theory is to account for the P-wave pairing in neutron superfluid.

\end{document}